\documentclass[a4paper]{article}

\usepackage{INTERSPEECH2015}

\usepackage{graphicx}
\usepackage{amssymb,amsmath,bm}
\usepackage{textcomp}
\usepackage{url}

\usepackage{bigstrut}
\usepackage{multirow}
\usepackage{array}
\usepackage{hyperref}
\usepackage{tabularx}

\ninept

\title{Multi-objective Learning and Mask-based Post-processing for Deep Neural Network based Speech Enhancement}

\makeatletter
\def\name#1{\gdef\@name{#1\\}}
\makeatother \name{{\em Yong Xu$^{1*}$\thanks{$^*$ This work is done while Yong Xu was visiting Georgia Tech in 2014-2015.}, Jun Du$^1$, Zhen Huang$^2$, Li-Rong Dai$^1$, Chin-Hui Lee$^2$}}

\address{$^1$National Engineering Laboratory for Speech and Language Information Processing,\\
	University of Science and Technology of China, China \\
	$^2$School of Electrical and Computer Engineering, Georgia Institute of Technology, USA \\
   {\small \tt xuyong62@mail.ustc.edu.cn, jundu@ustc.edu.cn, chl@ece.gatech.edu}
}

\begin{document}

  \maketitle
  \begin{abstract}
    We propose a multi-objective framework to learn both secondary targets not directly related to the intended task of speech enhancement (SE) and the primary target of the clean log-power spectra (LPS) features to be used directly for constructing the enhanced speech signals. In deep neural network (DNN) based SE we introduce an auxiliary structure to learn secondary continuous features, such as mel-frequency cepstral coefficients (MFCCs), and categorical information, such as the ideal binary mask (IBM), and integrate it into the original DNN architecture for joint optimization of all the parameters. This joint estimation scheme imposes additional constraints not available in the direct prediction of LPS, and potentially improves the learning of the primary target. Furthermore, the learned secondary information as a byproduct can be used for other purposes, e.g., the IBM-based post-processing in this work.
    A series of experiments show that joint LPS and MFCC learning improves the SE performance, and IBM-based post-processing further enhances listening quality of the reconstructed speech.


  \end{abstract}
  \noindent{\bf Index Terms}: speech enhancement, deep neural network, minimum mean square error, multi-objective learning, binary mask

  \section{Introduction}

    Classical speech enhancement (SE) approaches, such as spectral subtraction \cite{boll1979suppression}, MMSE-based spectral amplitude estimator \cite{ephraim1984speech, ephraim1985speech} and optimally modified log-MMSE estimator \cite{cohen2001speech, cohen2003noise}, are considered as unsupervised techniques having been studied extensively for several decades. Based on key assumptions for the interactions between speech and noise, the tremendous progress has been made for those techniques in the past. However some issues, such as fast changing noise (e.g., \textit{machine gun} \cite{varga1993assessment}) and negative spectrum estimation, still need to be addressed.

    On the other hand, supervised machine learning approaches have also been developed in recent years. They were shown to generate enhanced speech with good qualities \cite{mohammadiha2013supervised}. Non-negative matrix factorization (NMF) based speech enhancement \cite{mohammadiha2013supervised, wilson2008regularized} was one notable example in which speech and noise basis models were learned separately from training speech and noise databases. Then the clean speech could be decomposed given the noisy speech. However, speech and noise are assumed uncorrelated and it limited the quality of the enhanced speech signals. Following recent successes in deep learning based speech processing \cite{hinton2012deep, dahl2012context, zhang2013denoising} we have recently proposed a deep neural network (DNN) based speech enhancement framework \cite{xu2015regression, xu2014experimental, yong2014is} in which DNN was regarded as a regression model to predict the clean log-power spectra (LPS) features \cite{du2008speech} from noisy LPS features. DNN also acts as a mapping function to learn the relationship between clean and noisy speech features without imposing any assumption. Similar DNN-based speech denoising methods were also proposed in \cite{lu2013speech, xia2014wiener}. In \cite{huang2014deep, liu2014experiments}, DNN-based method was demonstrated to be better than the NMF-based methods in speech separation. In DNN-based speech enhancement, the minimum mean square error (MMSE) between the target features and the predicted features was always used as the objective function. It is difficult to design a better cost function to directly optimize the DNN model, especially for features that are correlated. In \cite{liu2014experiments} it was shown that other cost functions, such as the Kullback Leibler divergence \cite{kullback1997information} or the Itakura-Saito divergence \cite{itakura1968analysis}, all performed worse than the MMSE.

    In this paper, a multi-objective learning framework is proposed to optimize a joint objective function, encompassing errors not only for the primary clean LPS features but also errors in secondary targets for continuous features, such as MFCC, and for categorical information, such as ideal binary mask (IBM) \cite{wang2014training}. This joint optimization of different but related targets can potentially improve the DNN prediction performance of the primary target LPS which is then used to reconstruct the enhanced waveform. In the LPS domain, the target values of different frequency bins were predicted independently without any correlation constraint, and some knowledge in auditory perception \cite{wang2006computational} is not easily utilized. Nonetheless in the MFCC domain, mel-filtering is first applied and the correlation of each channel is represented in the MFCC coefficients. Furthermore, IBM is the most important concept in the computational auditory scene analysis (CASA) \cite{wang2006computational}. IBM which represents the noise-dominant or speech-dominant meta information can also improve DNN training and the estimated IBM could further be used for post-processing. Finally, MFCC and IBM can be combined together to help predict the target clean LPS features.

    In our SE experiments, we find that learning MFCC and/or IBM as secondary tasks provides improvements to DNN-based speech enhancement. Furthermore, IBM-based post-processing also gives an additional 1.5 dB improvement of segmental signal-to-noise ratio (SSNR) \cite{du2008speech}.


  \section{Multi-objective Learning for DNN-based Speech Enhancement} \label{sec:multitask_dnn}

    In \cite{xu2015regression, xu2014experimental}, DNN is adopted as a mapping function to predict the clean LPS features from the noisy LPS features. The relationship between the clean and noisy speech features can be well learned because nearly no assumptions were imposed during the training process. However, other DNN-based methods, such as binary or soft mask \cite{wang2013towards, narayanan2013ideal} based speech enhancement, assume that speech and noise are independent \cite{xu2015regression} at each time-frequency (T-F) unit.

    Normalized MMSE is used to update the DNN weights,
    \begin{equation}
    Er=\frac{1}{N}\sum_{n=1}^{N}\frac{\|\hat{\textbf{X}}_{n} (\textbf{Y}{_{n\pm{\tau}}},\textbf{W}, \textbf{b})-\textbf{X}_{n}\|_{2}^2}{\|\textbf{X}_{n}\|_{2}^2}.
    \label{DNNerrors}
    \end{equation}
    \noindent
    where $Er$ is the normalized mean squared error and it can also be treated as the reciprocal of signal-to-noise ratio (SNR). This normalized squared error always reduces the distribution diversity of the clean training data and makes DNN training more stable. It should be noted that all the input and output features are normalized with a global mean and variance of the noisy training data. Hence, $\hat{\textbf{X}}_{n}$ and $\textbf{X}_{n}$ denote the estimated and clean normalized LPS at sample index $n$, respectively, with $N$ representing the mini-batch size, $\textbf{Y}{_{n\pm\tau}} $ being the noisy LPS feature vector where the window size of the context is $2*\tau+1$, with $(\textbf{W}, \textbf{b})$ denoting the weight and bias parameters to be learned.

    In this study, multi-objective learning is proposed to jointly predict the primary LPS features together with other secondary continuous features, such as MFCC, or/and some discrete category information, such as IBM, to enhance DNN learning as follows,
        \begin{multline}
        	Er=\frac{1}{N}\sum_{n=1}^{N}\frac{\|\hat{\textbf{X}}_{n} (\textbf{Y}{_{n\pm\tau}},\textbf{Y}{_{n\pm\tau}^{\text{cont}}},\textbf{W}, \textbf{b})-\textbf{X}_{n}\|_{2}^2}{\|\textbf{X}_{n}\|_{2}^2} + \\
        	\alpha*\frac{1}{N}\sum_{n=1}^{N}\frac{\|\hat{\textbf{X}}_{n}^{\text{cont}} (\textbf{Y}{_{n\pm\tau}},\textbf{Y}{_{n\pm\tau}^{\text{cont}}},\textbf{W}, \textbf{b})-\textbf{X}_{n}^{\text{cont}}\|_{2}^2}{\|\textbf{X}_{n}^{\text{cont}}\|_{2}^2} + \\
        	        \beta*\frac{1}{N}\sum_{n=1}^{N}{\|\hat{\textbf{X}}_{n}^{\text{cate}} (\textbf{Y}{_{n\pm\tau}},\textbf{Y}{_{n\pm\tau}^{\text{cont}}},\textbf{W}, \textbf{b})-\textbf{X}_{n}^{\text{cate}}\|_{2}^2}.
        	\label{jointMFCC-ibm}
        \end{multline}
        \noindent
              \begin{figure}[t]
              	\centering
              	\includegraphics[width=2.1in]{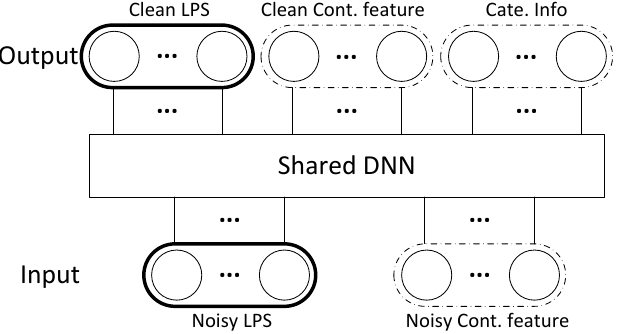}
              	\caption{{\it The structure of the multi-objective learning.}}
              	\label{fig:multi-learn-structure}
              \end{figure}
        where $\hat{\textbf{X}}^{\text{cont}}$ and $\textbf{X}^{\text{cont}}$ denote the estimated and clean continuous features (also normalized), respectively. $\textbf{Y}{^{\text{cont}}}$ represents the second noisy continuous feature. $\hat{\textbf{X}}^{\text{cate}}$ and $\textbf{X}^{\text{cate}}$ denote the estimated and target meta category information, respectively. $\alpha$ and $\beta$ are the weighting coefficients of this two other error parts, respectively. Unlike linear continuous features, meta information just has binary values, which makes the normalization not necessary for squared error related with the category part. Fig. \ref{fig:multi-learn-structure} presented the structure of the proposed multi-objective learning. In fact, it was similar to the multi-task learning \cite{caruna1993multitask}, but different from the multi-task learning in recent DNN-based speech recognition \cite{seltzer2013multi, Huang2015is} with only one input feature type. The prediction for the secondary continuous feature should be complementary with the prediction for the primary LPS  using the shared DNN. The learning for the category information with linear activation function should also promote the prediction of clean LPS. Overall, multi-objective learning can improve the generalization capability of DNN for the clean LPS estimation.

    \subsection{Joint Prediction of LPS with MFCC} \label{sec:multitask_dnn_jointmfcc}
    MFCC is one of the most popular speech features used in speech recognition \cite{vergin1999generalized}, speaker recognition \cite{murty2006combining} and music modeling \cite{logan2000mel}. Mel-filtering is applied to make it consistent with human auditory perception. However there is so far no prior auditory knowledge adopted in the LPS domain except for the log-compression. We believe the clean LPS features would be better predicted with a MFCC constraint imposed at the output layer. Furthermore, the discrete cosine transformation (DCT) \cite{ahmed1974discrete} operation in MFCC can incorporate the correlation information of different channels into each MFCC coefficient. We therefore expect correlated and consistent distortion across different frequency bins can be learned when predicting the clean LPS. Noted that DCT here is not performing dimension reduction which means the same dimensional MFCC features as the Mel-filter bank features are extracted.

    One similar work in \cite{wang2013exploring} showed that the concatenation of different input features could improve the performance of DNN-based speech separation. However the motivation of our work is multi-objective learning with a novel architecture in both input and output layers, which is totally different from the motivation of feature fusion in \cite{wang2013exploring}. It is expected that the enhancement of MFCC would be complimentary to the enhancement of LPS.

    \subsection{Joint Prediction of LPS with IBM}
    IBM \cite{wang2014training} is one type of category information often used to represent the  noise-dominant or speech-dominant nature at a certain T-F bin \cite{wang2006computational}. If the local SNR of a T-F bin is greater than a threshold, the IBM is set to one otherwise it is set to zero. Just like MFCC, IBM is also used as a constraint term in the joint objective function. IBM explicitly offers the additional speech presence information at T-F units. With this discriminative information, the speech components would be emphasized while reducing more noise components.

    In addition, the joint prediction of clean LPS with clean MFCC and IBM can be combined together. The noisy MFCC augmented in the input with the noisy LPS can also improve the IBM-based post-processing performance with an accurate IBM estimation to be discussed in the next section.
 \subsection{IBM-based Post-processing}
    The direct prediction of the clean LPS using DNN may lead to an overestimate or underestimate problem at some T-F units. The estimated IBM can be used for post-processing to simultaneously control the noise reduction level and speech distortion as follows,
    \begin{equation}
    {\hat{X}}_{n}^{\prime}(d)=\left\{
    \begin{array}{lcl}
     Y_n(d)      &      {\widehat{\text{IBM}}_n(d)\geq \gamma}\\
    \frac{(Y_n(d)+\hat{X}_n(d))}{2}    &    {\varepsilon < \widehat{\text{IBM}}_n(d) < \gamma}\\
     \hat{X}_n(d)     &    \text{otherwise}\\
    \end{array} \right.
    \label{eq:IBM-pp}
    \end{equation}
    where $\widehat{\text{IBM}}_n(d)$ denotes the estimated IBM at time frame $n$ and frequency bin {$d$}. Noted that the estimated IBM is close to the range $[0, 1]$. If the estimated IBM value is very large indicating that it has very high SNR at certain T-F unit, it is not necessary to perform noise reduction which can potentially result in the speech distortion. This is also the mask concept in \cite{wang2006computational}. If the estimated IBM has a medium value, the average value between the noisy LPS and the estimated LPS was used. Otherwise, the DNN predicted LPS was adopted. The proposed IBM post-processing scheme in Eq. (\ref{eq:IBM-pp}) is therefore different from \cite{wang2014training} where the estimated soft mask was used as a Wiener gain to perform speech enhancement. In contrast to adopting DNN to learn the mask \cite{wang2014training, wang2013towards} there is no independence assumption between speech and noise in our DNN based mapping strategy.

%

\section{Experimental Results and Analysis}\label{sec:experiments}
In \cite{xu2015regression, xu2014experimental}, all experiments were conducted on waveforms with 8kHz sample rate, in this work we extended it to 16kHz sample rate. 104 noise types were used in \cite{xu2015regression}, however, in this study 115 noise types including some musical noises were adopted to further improve the generalization capacity of DNN. These 115 noise types include 100 noise types recorded by G. Hu \cite{Hu100noises} and 15 home-made noise types\footnote[1]{The 115 noise types for training are N1-N17: Crowd noise; N18-N29: Machine noise; N30-N43: Alarm and siren; N44-N46: Traffic and car noise; N47-N55: Animal sound; N56-N69: Water sound; N70-N78: Wind; N79-N82: Bell; N83-N85: Cough; N86: Clap; N87: Snore; N88: Click; N88-N90: Laugh; N91-N92: Yawn; N93: Cry; N94: Shower; N95: Tooth brushing; N96-N97: Footsteps; N98: Door moving; N99-N100: Phone dialing; N101: AWGN; N102: Babble; N103-N105: Car; N106-N115: musical instruments. And all of them can be downloaded at {\href{http://home.ustc.edu.cn/~xuyong62/demo/115noises.html}{http://home.ustc.edu.cn/\~{}xuyong62/demo/115noises.html}}}. And the clean speech data is derived from the TIMIT corpus \cite{garofolo1988getting}. All 4620 utterances from the training set of the TIMIT database were corrupted with the abovementioned 115 noise types at six levels of SNR, i.e., 20dB, 15dB, 10dB, 5dB, 0dB, and -5dB, to build 80 hours multi-condition training set, consisting of pairs of clean and noisy speech utterances. The 192 utterances from the core test set of TIMIT database were used to construct the test set for each combination of noise types and SNR levels. As we only conduct the evaluation of unseen noise types in this paper, three other noise types, namely Buccaneer1, Destroyer engine and HF channel were adopted for testing. All of them are collected from the NOISEX-92 corpus \cite{varga1993assessment}. An improved version of OM-LSA \cite{cohen2003noise}, denoted as \textbf{LogMMSE}, was used for performance comparison with our DNN approach.

A short-time Fourier analysis was used to compute the DFT of each overlapping windowed frame. Then 257 dimensions LPS features \cite{du2008speech} were used to train DNNs. Segmental SNR (SSNR in dB) \cite{du2008speech}, perceptual evaluation of speech quality (PESQ) \cite{rix2001perceptual}, and short-time objective intelligibility (STOI) \cite{taal2011algorithm} were used to assess the quality and intelligibility of the enhanced speech. Frequency-dependent log-spectral distortion, defined as subtracting estimated LPS from clean LPS at each frequency bin, was also proposed to analyze the consistency of distortion across frequencies.
Rectified linear units (ReLU) \cite{dahl2013improving} was used as the activation function of DNN, and the DNN was initialized with random weights. Dropout \cite{srivastava2014dropout} and static noise aware training as in \cite{xu2015regression, seltzer2013investigation} were used to improve its generalization capacity for unseen noise environments. Mean and variance normalization was applied to the input and target feature vectors of the DNN. All DNN configurations were fixed at $L=3$ hidden layers, 2500 units at each hidden layer, and 7-frame input. The MFCC used in Section \ref{sec:multitask_dnn_jointmfcc} had 40 dimensions of static feature and one energy dimension using 40 Mel-filters. The empirical value of $\alpha$ and $\beta$ in Eq. (\ref{jointMFCC-ibm}) are set to 0.1 and 0.002, respectively. The empirical value of $\gamma$ and $\varepsilon$ in Eq. (\ref{eq:IBM-pp}) are set to 0.9 and 0.6, respectively.

\subsection{Joint Prediction of LPS and MFCC}
In Table \ref{tab:ibm}, average PESQ and SSNR comparison on the test set at different SNRs of the three unseen noise environments among: DNN baseline, MFCC augmented in the output (denoted as MFCC-o) and MFCC augmented in both the input and output (denoted as MFCC), were given. MFCC-o system consistently outperformed the DNN baseline in PESQ and SSNR which indicated that the simultaneous prediction of MFCC was beneficial for the estimation of clean LPS. Furthermore, the noisy MFCC was complementary with the noisy LPS in the input to improve the prediction of clean LPS. And the MFCC system got the best performance, such as the average PESQ improved from 2.508 to 2.664. The multi-task of MFCC enhancement and LPS enhancement shared the DNN weights and promoted each other.
     \begin{figure}[t]
     	\centering
     	\includegraphics[width=2.45in]{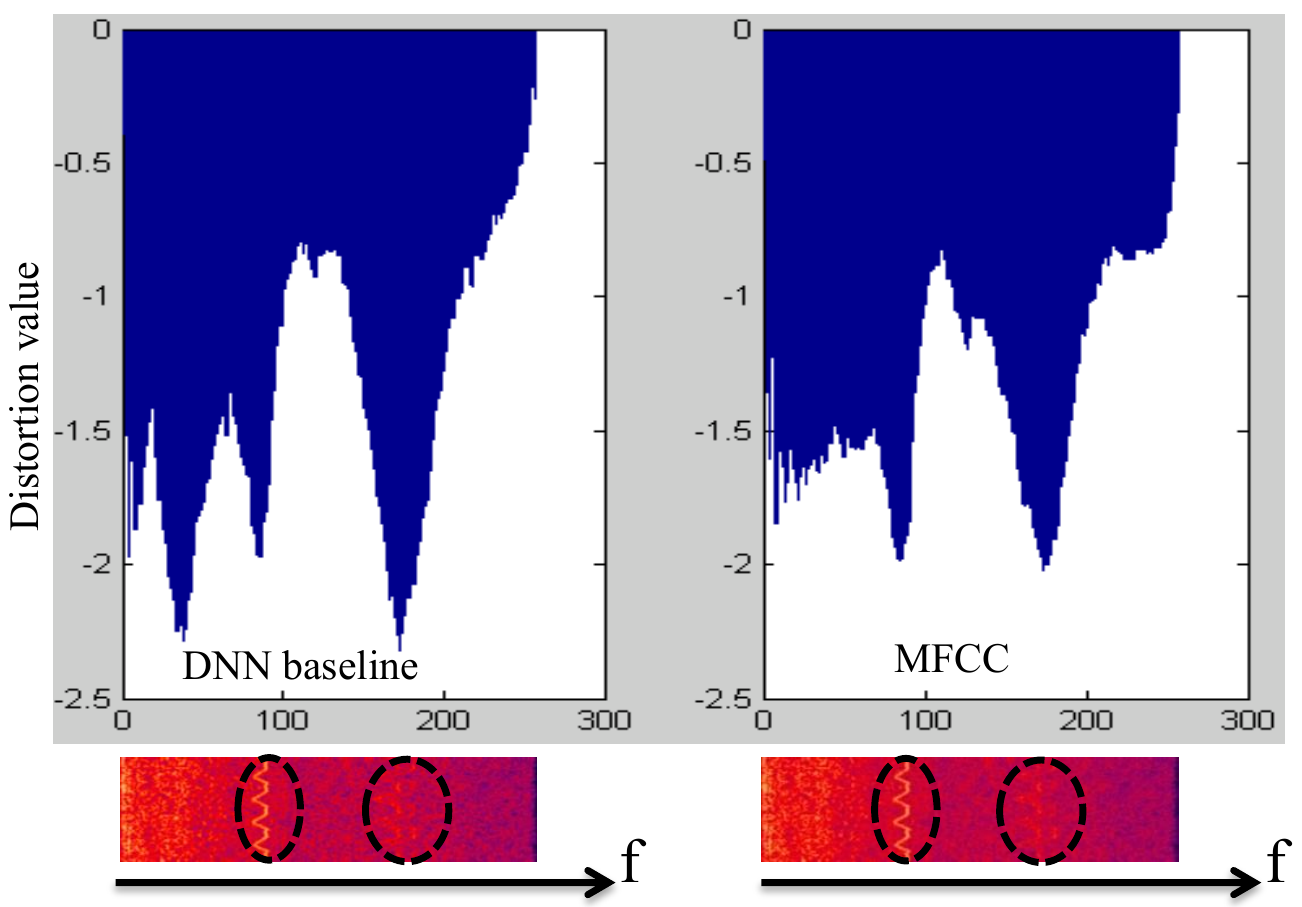}
     	\caption{{\it Frequency-dependent log-spectral distortion between the DNN baseline and MFCC systems calculated from 192 testing utterances at SNR=0dB corrupted by the Buccaneer1 noise (shown in the spectrogram above). And the x-axis is frquency.}}
     	\label{fig:bMFCC_corr}
     \end{figure}
The frequency-dependent log-spectral distortion between the DNN baseline and MFCC systems calculated from 192 testing utterances at SNR=0dB corrupted by the Buccaneer1 noise was also given in Fig. \ref{fig:bMFCC_corr}. The overall shape of this log-spectral distortion is determined by the noise type, such as here the Buccaneer1 noise has two continual and high energy parts at frequencies shown in the ellipses. But with the constraint of MFCC, the speech distortion at low frequencies where the most of speech info located was largely reduced and more consistent. This was because MFCC emphasized the info at low frequencies with the Mel-filtering.
  \begin{table*}[th]\scriptsize
  	\caption{\it Average PESQ and SSNR comparison on the test set at different SNRs of the three unseen noise environments, among: DNN baseline, MFCC-augmented output (denoted as MFCC-o), MFCC augmented in the input and output (denoted as MFCC), IBM augmented in the output of the DNN baseline without post-processing (denoted as IBM), IBM with post-processing (denoted as IBM+PP), MFCC and IBM without post-processing (denoted as MFCC+IBM) and MFCC and IBM with post-processing (denoted as MFCC+IBM+PP).}
  	\vspace{2mm}
  	\centerline{
  	\begin{tabular}{|c|c|c|c|c|c|c|c|c|c|c|c|c|c|c|}
  		\hline
  		\multicolumn{1}{|c|}{} & \multicolumn{2}{|c|}{Baseline} & \multicolumn{2}{|c|}{MFCC-o} & \multicolumn{2}{|c|}{MFCC} & \multicolumn{2}{|c|}{IBM} & \multicolumn{2}{|c|}{IBM+PP} & \multicolumn{2}{|c|}{MFCC+IBM} & \multicolumn{2}{c|}{MFCC+IBM+PP} \\
  		\hline
  		\multicolumn{1}{|c|}{SNR} & \multicolumn{1}{|c|}{PESQ} & \multicolumn{1}{|c|}{SSNR} & \multicolumn{1}{|c|}{PESQ} & \multicolumn{1}{|c|}{SSNR} & \multicolumn{1}{|c|}{PESQ} & \multicolumn{1}{|c|}{SSNR} & \multicolumn{1}{|c|}{PESQ} & \multicolumn{1}{|c|}{SSNR} & \multicolumn{1}{|c|}{PESQ} & \multicolumn{1}{|c|}{SSNR} & \multicolumn{1}{|c|}{PESQ} & \multicolumn{1}{|c|}{SSNR} & \multicolumn{1}{|c|}{PESQ} & \multicolumn{1}{|c|}{SSNR} \bigstrut[b]\\
  		\hline
  		\cline{2-15}\multicolumn{1}{|c|}{20} & 3.287 & 7.403 & 3.324 & 7.592 & 3.387 & 8.199 & 3.309 & 7.641 & 3.358 & 11.455 & 3.391 & 8.147 & \textbf{3.424} & \textbf{11.862} \bigstrut\\
  		\hline
  		\cline{2-15}\multicolumn{1}{|c|}{15} & 3.014 & 5.721 & 3.051 & 5.936 & 3.128 & 6.637 & 3.029 & 5.987 & 3.083 & 8.616 & 3.135 & 6.610 & \textbf{3.167} & \textbf{9.164} \bigstrut\\
  		\hline
  		\cline{2-15}\multicolumn{1}{|c|}{10} & 2.713 & 3.812 & 2.748 & 4.087 & 2.852 & 4.762 & 2.722 & 4.097 & 2.791 & 5.808 & 2.861 & 4.782 & \textbf{2.895} & \textbf{6.418} \bigstrut\\
  		\hline
  		\cline{2-15}\multicolumn{1}{|c|}{5} & 2.387 & 1.825 & 2.414 & 2.204 & 2.551 & 2.662 & 2.384 & 2.131 & 2.463 & 3.145 & 2.567 & 2.770 & \textbf{2.597} & \textbf{3.673} \bigstrut\\
  		\hline
  		\cline{2-15}\multicolumn{1}{|c|}{0} & 2.030 & -0.084 & 2.045 & 0.413 & 2.217 & 0.534 & 2.013 & 0.251 & 2.084 & 0.811 & 2.238 & 0.759 & \textbf{2.261} & \textbf{1.102} \bigstrut\\
  		\hline
  		\cline{2-15}\multicolumn{1}{|c|}{-5} & 1.617 & -1.693 & 1.624 & -1.171 & 1.847 & -1.433 & 1.603 & -1.314 & 1.679 & \textbf{-1.036} & 1.868 & -1.086 & \textbf{1.887} & -1.054 \bigstrut\\
  		\hline
  		\cline{2-15}\multicolumn{1}{|c|}{Ave} & 2.508 & 2.831 & 2.534 & 3.177 & 2.664 & 3.560 & 2.510 & 3.132 & 2.576 & 4.800 & 2.677 & 3.664 & \textbf{2.705} & \textbf{5.194} \bigstrut\\
  		\hline
  		\cline{2-15}\end{tabular}%
  	}
  	\label{tab:ibm}
  \end{table*}

\subsection{Joint Prediction of LPS and IBM with Post-processing}
Table \ref{tab:ibm} also presented the average PESQ and SSNR comparison for joint prediction of LPS and IBM on the test set at different SNRs of the three unseen noise environments.
With the IBM constraint in the output, better average PESQ and SSNR performance could be obtained compared with the DNN baseline, especially in SSNR which improved from -0.084 dB to 0.251 dB at SNR=0dB. Moreover, the IBM-based post-processing can obtain large PESQ and SSNR improvements, especially at high SNRs, e.g., SSNR improved from 7.641 dB to 11.455 dB at SNR=20dB which implies that the baseline DNN might hurt the speech components due to under-estimation, especially at the T-F units with high SNRs. Hence, IBM-based post-processing is crucial in achieving less speech distortion. This also conformed the mask concept in \cite{wang2006computational} that it was not necessary to reduce noise when the speech energy is much larger than the noise energy at the certain T-F unit.
In addition, IBM could be combined with MFCC. Compared with the performance of MFCC system, the combined system (MFCC+IBM in Table \ref{tab:ibm}) gave slightly better results at all SNR levels. For example SSNR was improved from -1.433 dB to -1.086 dB at SNR=-5dB. Finally, the average SSNR of the best MFCC+IBM+PP system was improved from 3.664 dB to 5.194 dB.

\subsection{Overall Performance Comparison}
        \begin{figure}[t]
        	\centering
        	\includegraphics[width=\linewidth]{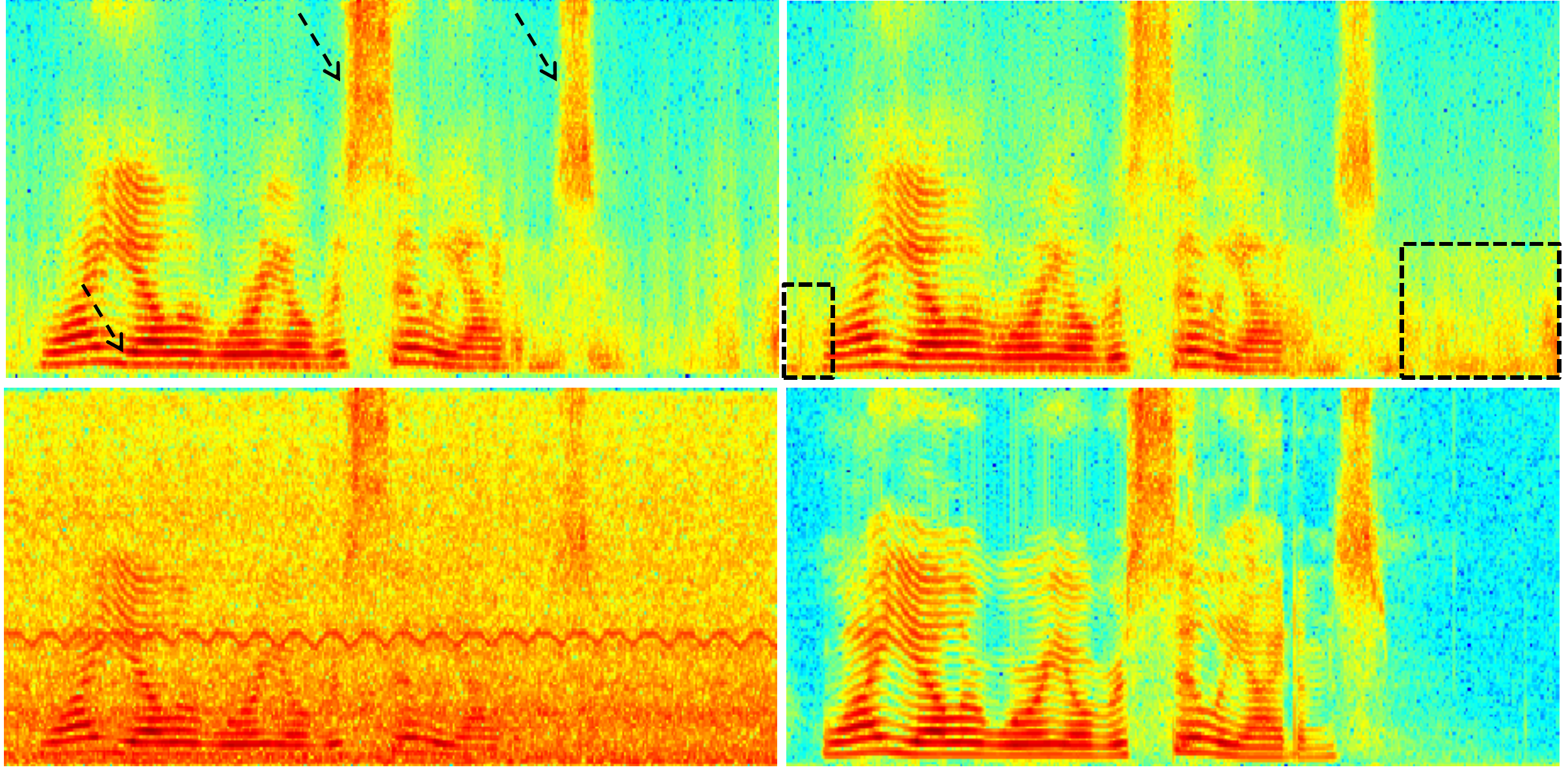}
        	\caption{{\it Comparison of four spectrograms of a 16kHz TIMIT utterance corrupted by Buccaneer1 noise at SNR=5dB: proposed DNN (upper left, PESQ=2.815), DNN baseline (upper right, PESQ=2.585), Noisy (bottom left, PESQ=1.591) and clean speech (bottom right, PESQ=4.5).}}
        	\label{fig:timit-16kHz}
        \end{figure}

  PESQ and STOI are often adopted to represent the objective quality and intelligibility of the enhanced speech, respectively. And STOI is often more meaningful at lower SNRs. An overall PESQ and STOI comparison of different SE techniques discussed in this study on the test set at different SNRs of the three unseen noise environments is displayed in Table \ref{tab:overall}. Compared with the noisy speech results, LogMMSE could yield 0.418 PESQ improvement while only 0.011 STOI improvement on average. The DNN baseline improved the LogMMSE with an average STOI from 0.801 to 0.845 across six SNRs. Our proposed MFCC+IBM+PP system overwhelms LogMMSE at all SNRs, especially at low SNRs, e.g., 0.163 STOI improvement and 0.626 PESQ improvement at SNR=-5dB.
          \begin{table}[th]\scriptsize
          	\caption{\it Average PESQ and STOI comparison on the test set at different SNRs of the three unseen noise environments, among: Noisy, LogMMSE \cite{cohen2003noise}, DNN baseline and the proposed MFCC+IBM+PP in Table \ref{tab:ibm} (denoted as Proposed).}
          	\vspace{2mm}
          	\centerline{
          		\begin{tabular}{|c|c|c|c|c|c|c|c|c|}
          			\hline
          			\multicolumn{1}{|c|}{} & \multicolumn{2}{|c|}{Noisy} & \multicolumn{2}{|c|}{LogMMSE} & \multicolumn{2}{|c|}{DNN Baseline} & \multicolumn{2}{|c|}{Proposed DNN} \\
          			\hline
          			\multicolumn{1}{|c|}{SNR} & \multicolumn{1}{|c|}{PESQ} & \multicolumn{1}{|c|}{STOI} & \multicolumn{1}{|c|}{PESQ} & \multicolumn{1}{|c|}{STOI} & \multicolumn{1}{|c|}{PESQ} & \multicolumn{1}{|c|}{STOI} & \multicolumn{1}{|c|}{PESQ} & \multicolumn{1}{|c|}{STOI} \bigstrut[b]\\
          			\hline
          			\cline{2-9}20 & 2.834 & 0.971 & 3.349 & 0.975 & 3.287 & 0.963 & \textbf{3.424} & \textbf{0.979} \bigstrut\\
          			\hline
          			\cline{2-9}15 & 2.481 & 0.934 & 3.049 & 0.945 & 3.014 & 0.944 & \textbf{3.167} & \textbf{0.960} \bigstrut\\
          			\hline
          			\cline{2-9}10 & 2.133 & 0.868 & 2.711 & 0.890 & 2.713 & 0.908 & \textbf{2.895} & \textbf{0.928} \bigstrut\\
          			\hline
          			\cline{2-9}5  & 1.793 & 0.772 & 2.299 & 0.800 & 2.387 & 0.849 & \textbf{2.597} & \textbf{0.876} \bigstrut\\
          			\hline
          			\cline{2-9}0  & 1.482 & 0.656 & 1.798 & 0.669 & 2.030 & 0.762 & \textbf{2.261} & \textbf{0.796} \bigstrut\\
          			\hline
          			\cline{2-9}-5 & 1.235 & 0.541 & 1.261 & 0.525 & 1.617 & 0.645 & \textbf{1.887} & \textbf{0.688} \bigstrut\\
          			\hline
          			\cline{2-9}Ave   & 1.993 & 0.790 & 2.411 & 0.801 & 2.508 & 0.845 & \textbf{2.705} & \textbf{0.871} \bigstrut\\
          			\hline
          			\cline{2-9}\end{tabular}%
          	}
          	\label{tab:overall}
          \end{table}
  Fig. \ref{fig:timit-16kHz} presented spectrograms of an utterance. The non-stationary noise was successfully reduced in the DNN-enhanced spectrum, while LogMMSE could not well track the non-stationary Buccaneer1 noise (its spectrogram can be seen at the demo website\footnote[2]{\href{http://home.ustc.edu.cn/~xuyong62/demo/IS15.html}{http://home.ustc.edu.cn/\~{}xuyong62/demo/IS15.html}}).  Compared with the baseline DNN-enhanced spectrogram, the improved DNN can enhance the speech with less speech distortion shown in the three dashed arrow areas, especially at the consonant portions which are similar to noise. Furthermore the improved DNN can also reduce noise shown in the rectangle highlight segments. More enhanced waveforms of real-world noisy speech can also refer to the website.

  \section{Conclusion}\label{sec:conclusion}
In this paper, multi-objective learning is proposed to improve DNN training for speech enhancement. Adding constraints from features like MFCC or IBM in the objective function is shown to obtain more accurate estimation of clean LPS. MFCC can make the log-spectral distortion more consistent across low frequencies; IBM can explicitly represent the speech presence information at T-F units, so higher SSNR could be obtained. Furthermore, the estimated IBM can be adopted to do post-processing to alleviate the over-estimate or under-estimate problems in regression-based DNN. And IBM-based post-processing was crucial to reduce speech distortion, especially at high SNR T-F units. Compared with DNN baseline, about 0.2 PESQ and 0.03 STOI improvements were obtained on average. In the future, other continuous features and meta information will be further explored.

\section{Acknowledgement}
This work was partially supported by the National Nature Science Foundation of China (Grant Nos. 61273264 \& 61305002).

  \newpage
  \eightpt
  \bibliographystyle{IEEEtran}

  \bibliography{mybib}


\end{document}